\documentclass[dvips]{acta}
\usepackage{supertabular,lscape,epsfig}
\usepackage{amssymb}
\usepackage{amsmath}
\DeclareSymbolFont{ppa}{OT1}{ppl}{m}{it}
\DeclareMathSymbol{\vv}{\mathalpha}{ppa}{'166}

\newfont{\hb}{rphvb at 10pt}
\newfont{\hbo}{rphvbo at 10pt}
\newfont{\bitt}{rptmbi at 12pt}
\newfont{\bits}{rptmbi at 11pt}

\SetPages{297}{311}

\SetVol{62}{2012}

\begin{document}

\newcommand{\TabApp}[2]{\begin{center}\parbox[t]{#1}{\centerline{
  {\bf Appendix}}
  \vskip2mm
  \centerline{\small {\spaceskip 2pt plus 1pt minus 1pt T a b l e}
  \refstepcounter{table}\thetable}
  \vskip2mm
  \centerline{\footnotesize #2}}
  \vskip3mm
\end{center}}

\newcommand{\TabCapp}[2]{\begin{center}\parbox[t]{#1}{\centerline{
  \small {\spaceskip 2pt plus 1pt minus 1pt T a b l e}
  \refstepcounter{table}\thetable}
  \vskip2mm
  \centerline{\footnotesize #2}}
  \vskip3mm
\end{center}}

\newcommand{\TTabCap}[3]{\begin{center}\parbox[t]{#1}{\centerline{
  \small {\spaceskip 2pt plus 1pt minus 1pt T a b l e}
  \refstepcounter{table}\thetable}
  \vskip2mm
  \centerline{\footnotesize #2}
  \centerline{\footnotesize #3}}
  \vskip1mm
\end{center}}

\newcommand{\MakeTableApp}[4]{\begin{table}[p]\TabApp{#2}{#3}
  \begin{center} \TableFont \begin{tabular}{#1} #4 
  \end{tabular}\end{center}\end{table}}

\newcommand{\MakeTableSepp}[4]{\begin{table}[p]\TabCapp{#2}{#3}
  \begin{center} \TableFont \begin{tabular}{#1} #4 
  \end{tabular}\end{center}\end{table}}

\newcommand{\MakeTableee}[4]{\begin{table}[htb]\TabCapp{#2}{#3}
  \begin{center} \TableFont \begin{tabular}{#1} #4
  \end{tabular}\end{center}\end{table}}

\newcommand{\MakeTablee}[5]{\begin{table}[htb]\TTabCap{#2}{#3}{#4}
  \begin{center} \TableFont \begin{tabular}{#1} #5 
  \end{tabular}\end{center}\end{table}}

\newfont{\bb}{ptmbi8t at 12pt}
\newfont{\bbb}{cmbxti10}
\newfont{\bbbb}{cmbxti10 at 9pt}
\newcommand{\uprule}{\rule{0pt}{2.5ex}}
\newcommand{\douprule}{\rule[-2ex]{0pt}{4.5ex}}
\newcommand{\dorule}{\rule[-2ex]{0pt}{2ex}}
\begin{Titlepage}
\Title{Variability of the Ap Star HD~9996} 
\Author{V.\,D.~~B~y~c~h~k~o~v$^1$,~~ L.\,V.~~B~y~c~h~k~o~v~a$^1$,~~ J.~~M~a~d~e~j$^2$~~
and~~ A.\,V.~~S~h~a~t~i~l~o~v$^1$}
{$^1$Special Astrophysical Observatory RAS, Niznyi Arkhyz, Russia\\
e-mail: vbych@sao.ru\\
$^2$Warsaw University Observatory, Al. Ujazdowskie 4, 00-478 Warszawa, Poland \\
e-mail: jm@astrouw.edu.pl}
\Received{July 16, 2012}
\end{Titlepage}

\Abstract{
We present here new measurements of the longitudinal magnetic field ($B_e$)
in the binary system HD~9996, where the primary companion is an Ap star.
Series of 63 $B_e$ observations was obtained in years 1994--2011 with the
1-m optical telescope of the Special Astrophysical Observatory (Russia).
New magnetic data allowed us to refine the long-term magnetic period of
HD~9996 to $P_{\rm mag}=21.8$ years. Compilation of archival photometric
data showed the existence of long-term variations of HD~9996 in time scale
of 22--23~yr consistent with $P_{\rm mag}$. We identify $P_{\rm mag}$ with
the precession period of the primary Ap star and summarize search for a
short-term rotational period and the rotational line broadening in this
star.}{Stars: chemically peculiar -- Stars: individual: HD~9996 -- Stars:
magnetic fields}

\Section{Introduction}
The star HD~9996~(HR465,~GY~And), spectral type B9p CrEuSi belongs to the
subclass of long-period magnetic Ap stars. Compilation of early photometric
measurements for HD~9996 allowed one to constrain the corresponding period
$P$ in the range $7750<P_{\rm ph}<8550$~days, or $21.2<P_{\rm
ph}<23.4$~years (Pyper and Adelman 1986). First measurements of the
effective (longitudinal) magnetic field $B_e$ in HD~9996 were initiated by
Babcock (1958) who used the photographic method for recording of
magnetically split spectral lines. Further measurements of $B_e$ were
published by Preston and Wolff (1970) and Scholz (1978, 1983), who used the
same method.

Our new magnetic measurements of HD~9996 were the result of a large-scale
observational program for measuring magnetic fields in stars, which was
performed at the Special Astrophysical Observatory of the Russian Academy
of Science. The program used 6-m optical telescope and the hydrogen-line
magnetometer (Bychkov 2000). Series of $B_e$ observations for this star was
published by Bychkov \etal (1997), who improved the magnetic period and
obtained $P_{\rm mag}=7\,842$~days, or 21.5~yr. However, there existed a
large uncertainty regarding both the exact value of that period and the
character of the magnetic variations $B_e$ \vs time, which did not form a
simple sine-wave as was in the oblique-rotator model.

From that time two of us (VDB and LVB) continued monitoring of the magnetic
field strength $B_e$ of HD~9996. This paper presents results of this 18~yr
observing run.

\Section{Instrumentation and Data Processing}
HD~9996 is a relatively bright object with the apparent visual magnitude
$V=6.38$~mag. Therefore, magnetic monitoring of this star was performed
in coude focus of the 1-m reflector at SAO, equipped with the CEGS
spectrometer and analyzer of circular polarization (\cf Bychkov 2008).

In this research we obtained high quality CCD spectra of the star with
$R=45\,000$, ${\rm S/N}\approx40$ or even more, which were processed with
MIDAS software. Values of the longitudinal magnetic field $B_e$ were
determined with the standard procedure consisting of the following steps.

\begin{itemize}
\parskip=0pt \itemsep=1mm \setlength{\itemsep}{0.4mm}\setlength{\parindent}{0em} \setlength{\itemindent}{0em}
\item[1.]Computer code STARSP (Tsymbal 1996) generated synthetic spectrum of the
star, using the database VALD (Kupka \etal 1999). Then, we selected a set
of spectral lines suitable for further measurements.
\item[2.]Gaussians were fitted  by the least squares method to the
observed profiles both in clockwise (RCP) and counterclockwise (LCP)
circularly polarized light. We ignored lines which exhibited registration
defects due \eg to cosmic rays. Line splitting caused by the longitudinal
magnetic field was set to the separation of the weight centers of both
circularly polarized Gaussian profiles.
\item[3.]Intensity of the effective magnetic field was derived from the separation
of both polarized components of each line following the well-known relation
$$\lambda_L-\lambda_R=2\,g_{\rm eff}\,4.67\times 10^{-13}\,\lambda_0^2\,B_e\,,\eqno(1)$$
where the wavelength $\lambda_0$ was expressed in \AA\ and the magnetic
field intensity $B_e$ in~G.
\end{itemize}

The final value of $B_e$ was the arithmetic mean of the intensities taken
over all useful spectral lines in a given spectrum. We estimated the
standard deviation $\sigma$ of $B_e$ assuming normal distribution of all
errors.

Instrumental effects were taken into account following Bychkov, Romanenko
and Bychkova (2000). During each observing night we performed measurements
of the magnetic standard stars, $\alpha^2$~CVn and 53~Cam, as well as two
standard stars of ``null'' magnetic field, $\alpha$~CMi, $\alpha$~Boo, also
the Moon etc. Measurements of $B_e$ for ``standard'' stars were in a good
agreement with the data collected in the literature.

The above observational and calibration procedures and reduction of raw
data were also described in Bychkov, Bychkova and Madej (2006).

\Section{Observations}
In 2005 (JD 2453370.+) we collected 39 new $B_e$ observations of HD~9996 and 
determined the period of magnetic variability $P_{\rm mag}=7\,692$~days
(Bychkov \etal 2005).

\MakeTable{|l|r|r|c@{\hspace{1cm}}|l|r|r|}{12cm}{Magnetic field measurements $B_e$ of the Ap star HD~9996
obtained with the 1-m telescope SAO RAS}
{\cline{1-3}\cline{5-7}
\douprule JD2400000.+ & \multicolumn{1}{c|}{$B_e$} &
\multicolumn{1}{c|}{$\sigma_{B_e}$}
&&JD2400000.+ & \multicolumn{1}{c|}{$B_e$} & \multicolumn{1}{c|}{$\sigma_{B_e}$}\\
\cline{1-3}\cline{5-7}
  49649.378  & $-2231$ &   92  &&  54779.3493 & $+ 385$ &   84 \\
  50022.301  & $-1730$ &   82  &&  54781.3159 & $+ 609$ &   59 \\
  50022.347  & $-1800$ &   78  &&  54783.2618 & $+ 529$ &   65 \\
  50023.388  & $-1683$ &   69  &&  55080.4590 & $+ 145$ &  106 \\
  50064.282  & $-1831$ &   86  &&  55081.4097 & $+ 509$ &   39 \\
  51535.2986 & $- 146$ &   69  &&  55082.3777 & $+  23$ &   59 \\
  51536.2139 & $- 361$ &   59  &&  55083.4833 & $+ 216$ &   72 \\
  51890.251  & $+ 214$ &  124  &&  55142.3229 & $+ 966$ &   79 \\
  53273.569  & $+ 670$ &   71  &&  55164.2991 & $+ 162$ &   93 \\
  53275.622  & $+ 453$ &   99  &&  55494.3313 & $-  59$ &   52 \\
  53276.613  & $+ 552$ &  120  &&  55494.3736 & $-  60$ &   61 \\
  53278.602  & $+ 723$ &   90  &&  55495.3951 & $- 217$ &  104 \\
  53279.616  & $+ 843$ &   95  &&  55553.1340 & $- 140$ &   60 \\
  53626.484  & $+ 547$ &   90  &&  55553.1763 & $- 141$ &   59 \\
  53629.500  & $+ 707$ &   86  &&  55554.1263 & $- 180$ &   67 \\
  53632.401  & $+ 721$ &   99  &&  55554.1631 & $- 181$ &   61 \\
  53636.373  & $+ 583$ &  147  &&  55555.1187 & $-  81$ &   70 \\
  53637.348  & $+ 611$ &   84  &&  55555.1548 & $-  82$ &   58 \\
  53638.377  & $+ 740$ &  120  &&  55584.1423 & $- 107$ &   59 \\
  53665.342  & $+ 268$ &  118  &&  55584.1861 & $- 111$ &   73 \\
  53666.289  & $+ 382$ &  105  &&  55819.5076 & $- 467$ &   47 \\
  53667.297  & $+ 821$ &   84  &&  55819.5437 & $- 469$ &   53 \\
  53668.312  & $+ 488$ &  132  &&  55823.5305 & $- 401$ &  102 \\
  53692.259  & $+ 431$ &   92  &&  55823.5673 & $- 403$ &  130 \\
  53718.171  & $+ 566$ &  101  &&  55852.3381 & $- 484$ &   50 \\
  53719.197  & $+ 443$ &  120  &&  55852.3777 & $- 486$ &   43 \\
  53721.199  & $+ 317$ &  101  &&  55852.4138 & $- 488$ &   74 \\
  54112.126  & $+ 771$ &  111  &&  55881.3333 & $- 420$ &   42 \\
  54373.4118 & $+ 633$ &  101  &&  55881.4194 & $- 424$ &   49 \\
  54374.3722 & $+ 676$ &  122  &&  55911.2875 & $- 500$ &   95 \\
  54429.2590 & $+ 827$ &   74  &&  55911.3159 & $- 502$ &   78 \\
  54430.2500 & $+ 998$ &   71  &&             &        &      \\
\cline{1-3}\cline{5-7}}
Since the magnetic phase curve was still poorly constrained, the monitoring
project for HD~9996 was continued until now. Finally, we collected 63
measurements of $B_e$ during the last 18 years. All our observations of
$B_e$ are shown in Table~1. Thus the total span of observations of $B_e$ in
HD~9996 amounts to 62 years, or almost 3 full magnetic periods $P_{\rm
mag}$.

Fig.~1 presents the measured values of the effective magnetic field $B_e$
for HD~9996 as a function of time (Julian Day).
\begin{figure}[htb]
\includegraphics[width=12.5cm]{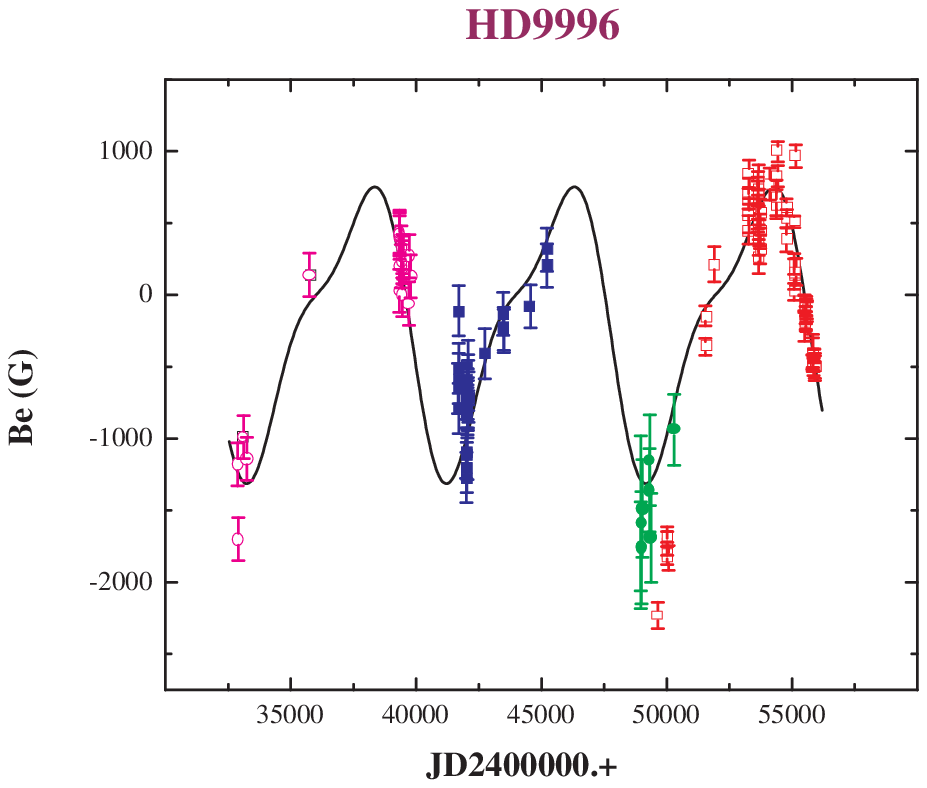}
\FigCap{Longitudinal magnetic field $B_e$ for HD~9996 \vs Julian Day in
the years 1946--2011. Observations are colored depending on the source:
magenta -- Babcock (1958), purple -- Preston and Wolff (1970), blue --
Scholz (1978, 1983), green -- Bychkov \etal (1997) 6-m telescope; red --
this paper (Table~1) 1-m telescope. Series of discrete $B_e$ points are
overlayed by the best fitted double sine phase curve.}
\end{figure}

We attempted to improve the magnetic period of HD~9996 using all the
available $B_e$ measurements, though the distribution of all $B_e$ points
\vs Julian Day is not well suited for the period analysis. We
obtained the best value $P_{\rm mag}=7961.8\pm22$ days, or 21.8 years. Zero
phase of $B_e$ variations was set to the time $T_0$ of minimum magnetic
field strength, $B_e{\rm (min)}$. The best fit yielded $T_0=2433240.7$~JD
(see the following sections).

The corresponding smooth curve approximating all available $B_e$ points was
defined by Eqs.~(2)--(3) and it is plotted in Figs.~1--2.

\Section{Magnetic Phase Curves}
\subsection{Long-Term Variations}
We approximated variations of the effective magnetic field of HD~9996,
$B_e$ \vs time $t$ by a double sine wave (Bychkov \etal 2005)
$$B_{ei}(\phi)=B_0+B_1\cos(\phi+z_1)+B_2\cos({2\phi}+z_2)\eqno(2)$$
where phase $\phi$
$$\phi=2\pi\frac{T_i-T_0}{P_{\rm mag}}.\eqno(3)$$

\begin{figure}[htb]
\vglue-5mm
\includegraphics[width=12cm]{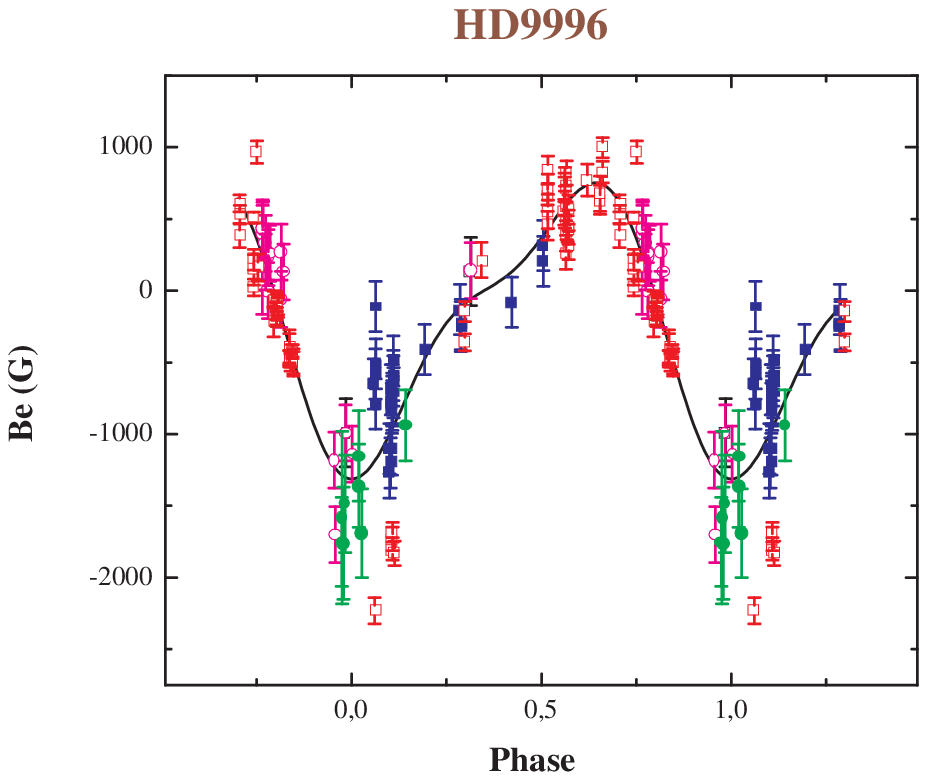}
\vglue-7mm
\FigCap{Magnetic phase curve $B_e(\phi)$ computed for the principal period 
$P_{\rm mag}=7961.8$ days $=21.8$~years. All the existing observations 
spanning 1946--2011 were taken into account here. The color coding
is the same as in Fig.~1.}
\end{figure}
Fig.~2 presents the magnetic phase curve $B_e(\phi)$ which was computed for
the period $P_{\rm mag}=7961.8$ days and the zero epoch
$T_0=2433240.7$~JD. Points $B_e$ on the curve are nonuniformly distributed
in phase $\phi$. Moreover, the points close to the phase $\phi\approx0$
are significantly scattered. Such a scatter suggests that the observed
variations of the longitudinal magnetic field $B_e$ in HD~9996 were caused
by superposition of a few processes with significantly differing periods.

One should note that the earliest magnetic field measurements $B_e$ were
performed with the photographic technique of low accuracy. The
corresponding points in Figs.~1--2 exhibit a rather large dispersion along
the vertical axis. Due to that, only our newest photoelectric $B_e$
measurements of HD~9996 from the last 18 years are the most significant and
were analyzed in this paper (\cf also Bychkov \etal 1997, 2005).

\MakeTable{|c|c|}{12.5cm}{Parameters of the long-term magnetic
variability of HD~9996. They define phase curve plotted in Fig.~3, which
was obtained only from our new observations presented in Table~1.}
{\hline
Parameter      & value   \\
\hline
$T_0$          & JD 2449478.4  \\
$P_{\rm mag}~~~~$  &  7961.8  days \\
$B_0$          &  $­415$    G \\
$B_1$          &   1311     G \\
$z_1$          &    0.985     \\
$B_2$          &   315      G \\
$z_2$          &   0.019      \\
\hline}

\begin{figure}[htb]
\vglue-11mm
\includegraphics[width=12cm]{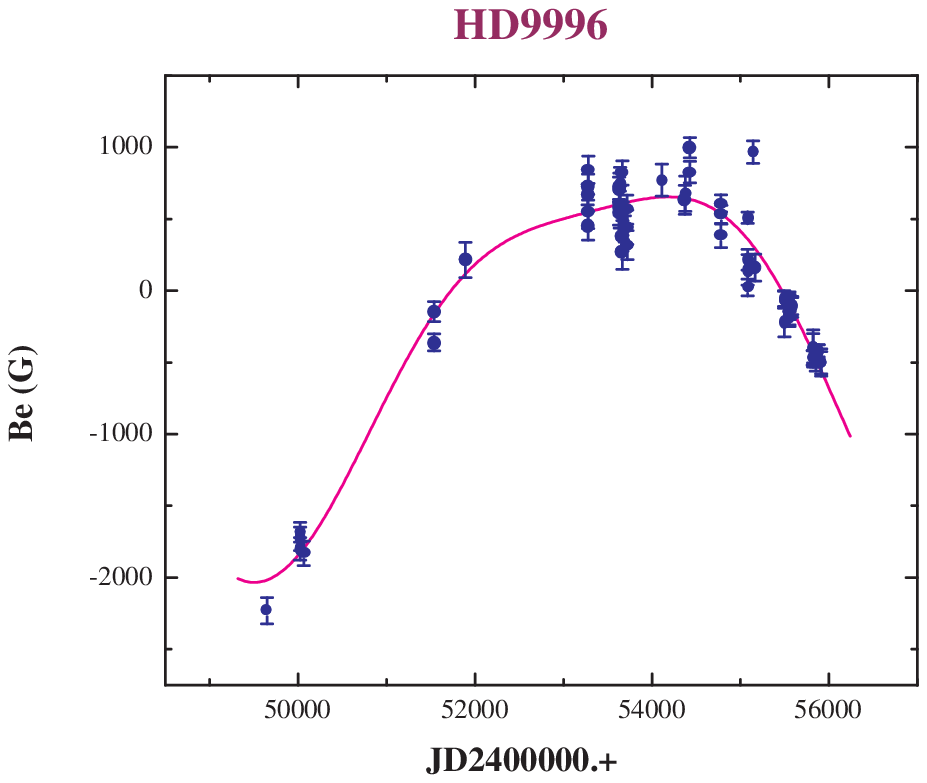}
\vglue-7mm
\FigCap{Magnetic field measurements obtained with the 1-m telescope
as a function of JD and magnetic phase curve, according to Eq.~(2) -- solid
line.}
\end{figure}
Fig.~3 shows the individual $B_e$ observations and error bars, which were
obtained at the Coude focus of the 1-m telescope at SAO RAS and the
instrumentation: 1­m telescope + GECS + ACP (with the subtraction of
instrumental polarization). Fig.~4 shows the magnetic phase curve
$B_e(\phi)$ derived only from our photoelectric measurements of the last 18
years. In such a manner we obtained a homogeneous series of $B_e$ points
of a fairly high accuracy, see Table~1. Parameters of the phase curve are
given in Table~2.
\begin{figure}[htb]
\vglue-16pt
\includegraphics[width=12.5cm]{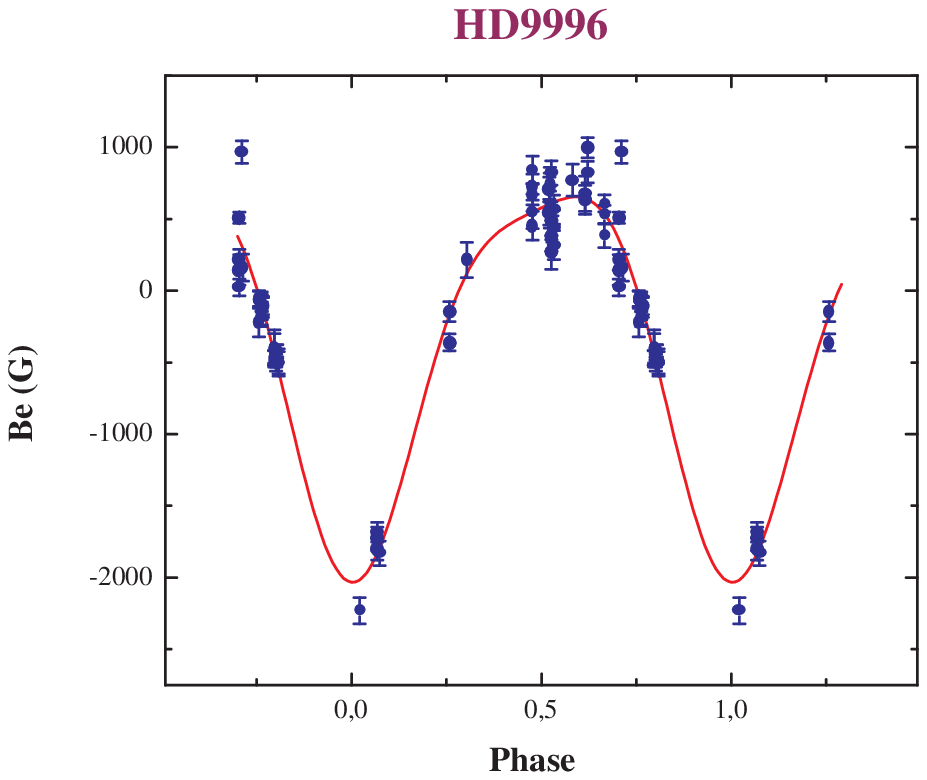}
\vglue-5mm
\FigCap{Magnetic phase curve $B_e(\phi)$ computed for the same period 
$P_{\rm mag}=7961.8$ days. The phase curve was derived only from our 
observations obtained at the 1-m telescope of SAO RAS.}
\end{figure}

\subsection{Search for Short-Term Variations}
Single-lined spectroscopic binary star HD~9996 is a rather wide pair with
the orbital period of $P_{\rm orb}=272.99$ days. Orbital elements of that
system are presented in Table~3, following Scholz (1978). Unfortunately,
errors of orbital elements were not given in his paper.
\MakeTable{|l|c|}{12.5cm}{Orbital elements of the binary system 
HD~9996 after Scholz (1978)}
{\hline
Parameter          & Value   \\
\hline
$e$                &  0.47 \\
$\omega$           &  $17\zdot\arcd7$ \\
$K$                &  11.3~km/s \\
$\gamma$           &  $-0.35$~km/s \\
$T_{\rm per}$      &  JD~2442048.03 \\
$a\sin i_{\rm orb}$& $37.4\times10^6$ km \\
\hline
\noalign{\vskip5pt}
\multicolumn{2}{p{4cm}}{Parameter $T_{\rm per}$ denotes the time of
periastron passage.}}

Carrier \etal (2002) determined the new orbital parameters for HD~9996,
with $P_{\rm orb}=272.88\pm0.20$ days, $T_{\rm per}={\rm JD}~2444492.34\pm
2.24$ and $e=0.532\pm0.023$. Both old and new periods $P_{\rm orb}$ and the
resulting orbital phases differ marginally, by less than the corresponding
errors.

In order to investigate the short-term $B_e$ variations we subtracted the
double-wave defined by Eq.~(2) from the 18 year series of our $B_e$
measurements. Long $P_{\rm mag}=7961.8~{\rm days}=21.8$ years period was
applied here. Spectral analysis of such a prewhitened $B_e$ time series
produced a white frequency spectrum with no features for frequencies $f\le
0.2$~day$^{-1}$. This result excluded the existence of periodic $B_e$
variations in HD~9996 with periods $P\ge5$ days and full amplitude
$\ge200$~G, see Fig.~5.
\begin{figure}[htb]
\vglue-11mm
\centerline{\includegraphics[width=10.7cm]{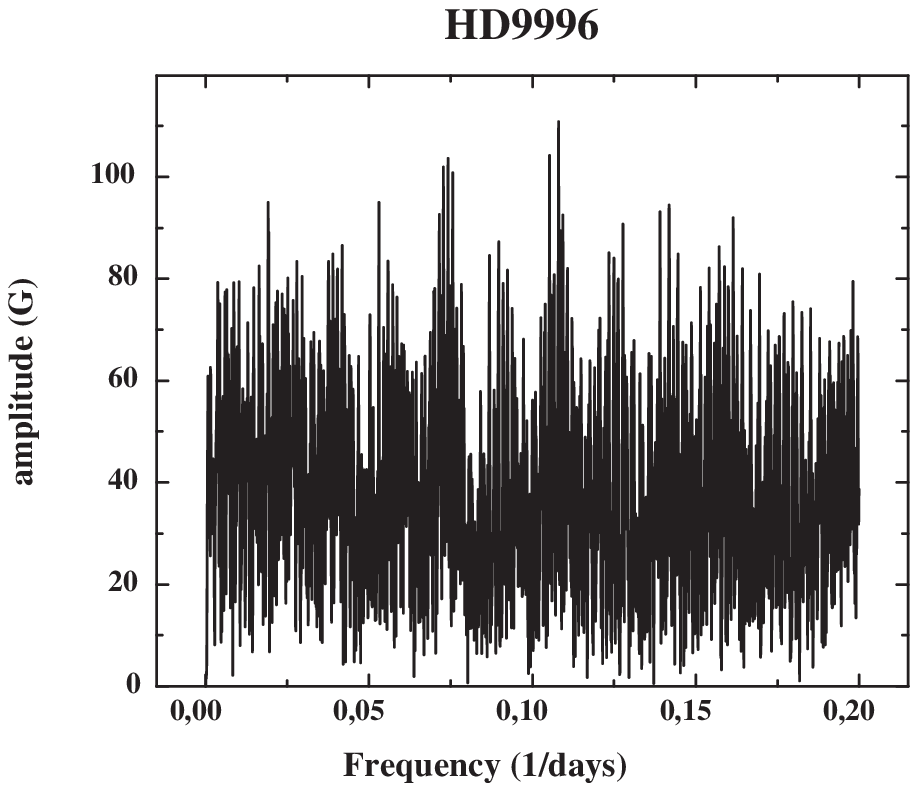}}
\vglue-3mm
\FigCap{Amplitude spectrum of the prewhitened time series of all our $B_e$
measurements (see Table~1).}
\end{figure}

In particular, there exist no detectable variations of the longitudinal
magnetic field $B_e$ with the orbital period $P_{\rm orb}=272.99$ exceeding
observational error, the latter of the order of 100--200~G. Such an
observation implies that the influence of the secondary on the primary Ap
star is rather weak due to a large separation. On the other hand, the
secondary on highly eccentric orbit still produces a periodic mechanical
impact on the Ap component at the moment of periastron passage.

Results of the spectral analysis of our $B_e$ time series was performed
following the method by Kurtz (1985) and with the original numerical code
supplied by the author. As can be seen from Fig.~5 there is no trace of the
period below the frequency $f\le0.2$~d$^{-1}$ with a half amplitude
bigger that $\approx100$~G.

\Section{Photometry}
\subsection{Long-Term Variations}
HD~9996 shows extremely slow magnetic, spectral and photometric variations,
which was noted by Preston and Wolff (1970). They found a slow decline of
{\it V} magnitude for HD~9996 by 0.1~mag, compared to {\it UBV}
observations by Abt and Golson (1962) and Stêpieñ (1968), which were made 8
years apart. Preston and Wolff (1970) concluded that the brightness of
HD~9996 varies with the period of 22--24 years.

The early three-color photometric observations of HD~9996 were also
published by Osawa and Hata (1962), but their results apparently came
unnoticed by Preston and Wolff in their paper.

Dumont and Le Borque (1983) appended new observations of HD~9996 which
originally were obtained in the Geneva photometric system and were
transformed later to the Johnson {\it V} and $B-V$ magnitudes. They
compiled all previous observations and eventually confirmed the possibility
of a 22--24 year period.

The newest {\it V} observations were compiled by Pyper and Adelman
(1986). Apparent luminosity of HD~9996 in these years varied in the range
$V{=}6.29{-}6.41$~mag.

\begin{figure}[htb]
\centerline{\includegraphics[width=13.1cm]{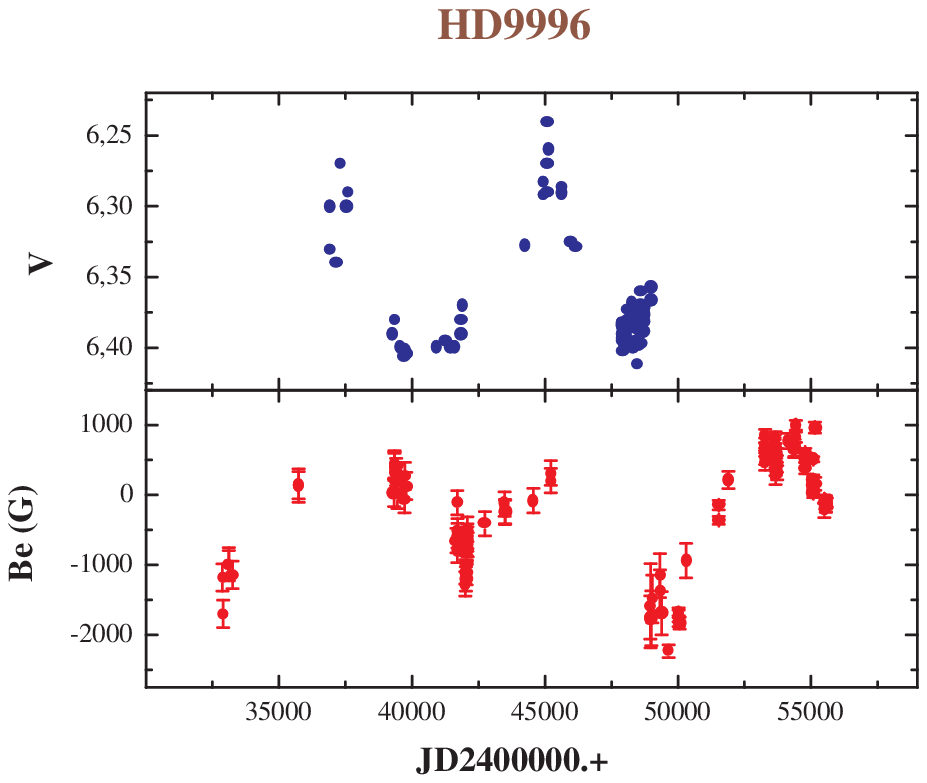}}
\FigCap{Plot of all available {\it V} magnitudes for HD~9996 \vs Julian Day
({\it upper panel}). Clump of points on the right side denotes $H_p$
magnitudes taken from the Hipparcos archive. {\it Lower panel} presents
determinations of the longitudinal magnetic field $B_e$.}
\end{figure}
Fig.~6 shows all the available {\it V} observations against Julian Day
(upper panel), including those obtained from the Hipparcos catalog (see
Section~5.3). For a comparison, the lower panel of Fig.~6 presents all the
existing magnetic $B_e$ measurements \vs JD.

\subsection{Short-Term Variations: Ground-based UBV observations}
Winzer (1974) claimed the existence of another photometric period in
HD~9996, $P_{\rm ph}=36.5$ days. This result was also quoted in the catalog
of periods of Ap stars by Catalano and Renson (1984). Photometry by Rakosch
and Fiedler (1978) gave much less conclusive results. Nevertheless, they
constrained the period to $P_{\rm ph}=35{-}40$ days, based on a rather
scarce set of their {\it B} and {\it V} magnitudes.

Rakosch and Fiedler (1984) constrained amplitudes of short-term light
variations to $\Delta U=0.012$~mag, $\Delta B=0.065$~mag and $\Delta
V=0.050$~mag, respectively. They presented the fine run of {\it U}
magnitudes \vs time, which suggested that the {\it U} luminosity was nearly
constant over about 100 days of observations, except for a small systematic
increase of {\it U}.

\subsection{Short-Term Variations: Hipparcos Photometry}
Hipparcos catalog contains time series of 90 $H_P$ magnitudes of HD~9996
measured during 3.5 years. Photon-counting tube device at the satellite
measured the brightness of stars in its own broad-band photometric system,
with the peak response at about 4500~\AA\ and a wide wing
redwards. Obviously, $H_P$ magnitude of a star can differ slightly from its
{\it V} brightness in the Johnson system.

The median magnitude of HD~9996 was equal to $H_P=6.3812$~mag at that time
with modulation of the amplitude $\Delta H_P=0.014$~mag and a period of
$P_{\rm ph}=39.76\pm0.02$ days. Therefore, Hipparcos photometry seemed to
confirm more uncertain results of earlier ground-based observations.

We reanalyzed time series of 93 original Hipparcos magnitudes of
HD~9996 (HIP 7651), which are available at
\vskip6pt
\centerline{\it http://www.rssd.esa.int/hipparcos\_scripts/HIPcatalogueSearch.pl?hipepId=7651}
\vskip6pt
\noindent
Amplitude spectrum of these observations shown in Fig.~7 does not show any
peak for periods $P\ge5$ days (frequency $f\le0.2$~days$^{-1}$) with full
amplitude higher than 0.012~mag. Therefore, we reject the period of 39.76
days claimed by the Hipparcos team in our analysis.

\begin{figure}[htb]
\includegraphics[width=12.5cm]{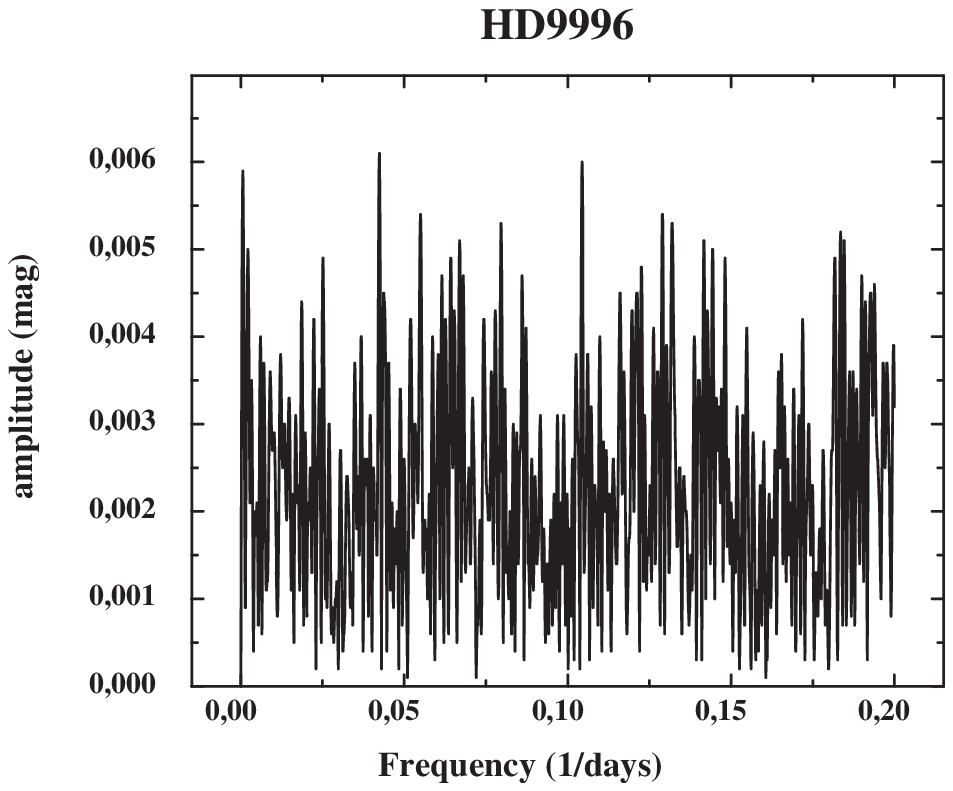}
\FigCap{Amplitude spectrum of Hipparcos $H_p$ magnitudes. No
trace of a periodic light variations is seen in this frequency range with
half-amplitude exceeding 0.006~mag.}
\end{figure}

\Section{Rate of Rotation}
We also measured half-widths of many spectral lines in our Zeeman spectra
to determine or constrain projected rotational velocity of the magnetic
star in HD~9996. We took into account lines in 20 spectra taken in over 17
years of observations, \ie during most of the observing
period. Measurements of the half-widths of the total 8723 narrow lines
showed, that HD~9996 rotates slowly and -- on the average -- line widths
are the same as in standard stars with $\vv_e\sin i=0$. Therefore, we can
only set an upper limit on the projected rotational velocity on the
equator
$$\vv_e\sin i\le8~{\rm km/s}\eqno(4)$$
and this value is just equal to  the average error of measurements.

The effective temperature of the primary Ap star was found in literature,
$T_{\rm eff}=10\,000$~K. This value implies that the radius equals to
$R=2.4$~\RS, which is valid for a main-sequence star. Preston (1971b)
presented a simple relation
$$\vv_e=\frac{50.61~R}{P_{\rm rot}}\eqno(5)$$
where $\vv_e$, $R$, and $P_{\rm rot}$ are expressed in km/s, solar radii and
days, respectively. Consequently, our extensive observations of half-widths
exclude short rotational periods of HD~9996 and set
$$P_{\rm rot}\ge15~{\rm days}.\eqno(6)$$

Our results agree with the early paper by Preston (1971a), who estimated
the projected equatorial velocity for HD~9996, $\vv_e \sin i \le 6 \> {\rm
km/s}$.  Resolution $R$ of our spectra did not allow us to constrain
further either $\vv_e\sin i$ or $P_{\rm rot}$.

We should note here that Carrier \etal (2002) published more stringent
estimate $\vv_e\sin i\le~2$ km/s, hence $P_{\rm rot}\sin i\ge60.7$~days.

\Section{Discussion}
\subsection{Photometry and Magnetic Field}
Spectroscopic binary star HD~9996 exhibits periodic magnetic and light
variations with at least one period. Long-term light and $B_e$ variations
were displayed in Fig.~6. They are apparently correlated and perhaps
proceed with the same period $P_{\rm mag}=21.8$~years.

Unfortunately, there exist no reliable V magnitudes of HD~9996 obtained
after the Hipparcos era. HD~9996 is a relatively bright star, therefore,
all recent robotic photometric surveys like TASS, WASP and ASAS3 produced
saturated images of this object and unreliable {\it V} determinations.

Hipparcos photometric data yielded the short-term period, $P_{\rm
ph}=39.76$ days. This is possibly is the rotational period, however, we did
not confirm its existence by an independent analysis of Hipparcos
observations.

\subsection{Precession and Tidal Interactions}
We attribute the period of $P_{\rm mag}=21.8$ years to the period of forced
precession of the rotational axis of the primary Ap star in the
gravitational field of the secondary. This type of precession in a binary
system was proposed and studied in detail by Shore and Adelman (1976) and
Lehmann (1987).

Further studies of the interaction between companion stars in HD~9996
require a model of precession and nutation of Ap star in this binary
system, taking into account exact value of its moment of inertia, the
latter taken from numerical models of stellar structure.

\subsection{Projected Equatorial Velocity}
Binary star HD~9996 possibly exhibits the following unique property. If the
period of precession in the system equals $P_{\rm mag}=21.8$~yr, then the
inclination angle $i$, hence $\vv_e\sin i$, must also oscillate with the
precession period. Detection of this effect in slowly rotating primary Ap
companion in HD~9996 requires use of optical spectra of high spectral
resolution for exact measurements of $\vv_e\sin i$ over many years.

Obviously the above effect of forced precession would be best observed when
the equator velocity $\vv_e$ was large enough or, consequently, when the
period of rotation is short. We failed even to measure a single $\vv_e\sin
i$ value in HD~9996 due to the slow rotation of the Ap companion and the
low resolution of our spectra ($R\approx25\,000$).

\subsection{Periods of HD~9996}
On the basis of the available observational data we propose the following
scenario. The observed effective magnetic field $B_e$ is the projected
field of the bright Ap primary star in the binary system. Long period of
magnetic variations $P_{\rm mag}=21.8$ years is just the precession period
of the primary star in the gravitational field of the secondary, which is
very faint in visual and remained unseen in the spectrum of the binary.

Light variations in the {\it V} Johnson filter are of the same time-scale
as the magnetic period of 21.8 yr equal by assumption to the precession
period. Note, that Rice (1988) identified a period of 21--22~yr with the
rotational period.

\section{Geometry of the Magnetic Field}
Surface properties of the primary Ap star in HD~9996 can be explained in
terms of the standard oblique-rotator model. Origin of its long-term
magnetic and light variations can be attributed to the precession of the
rotation axis in a gravitational field of the secondary, Both magnetic and
light variations must then proceed with the same period $P_{\rm
mag}=21.8$~yr, since both were caused by periodic variations of the aspect
angle. Note, that the variability period could not be determined with such
a high accuracy.

Possible short-term light variations with the period of ca.\ 40 days are
then caused by rotation of the Ap star. In such a scenario we must assume
that the magnetic axis is aligned to the stellar rotation axis. This is the
only configuration consistent with a global longitudinal field that is
constant as the star rotates. Such a configuration allows one to interpret
the observed 21.8 years field variability in terms of precession of the
star's rotation axis.

The requirement that the magnetic axis is parallel to the rotation axis
would be statistically not very likely, although Landstreet and Mathys
(2000) argued that slowly rotating stars tend to have the magnetic axis
tilted at a small angle with respect to the rotation axis.

Our results indicate however, that the scatter of the magnetic field
measurements at a given phase of the long period amounts to 200~G (see
Fig.~4). Apart from the measurement errors they may (at least partly)
result from modulation with the rotational period. In this case, both axes
may be inclined to each other, though we did not develop a numerical code
to compute models of $B_e$ variations of the oblique rotator.

Our periodogram in Fig.~7 excluded the existence of 40 days variations in
Hipparcos photometry with full amplitude bigger than 0.012~mag. The period
of 39.76 days obtained by Hipparcos team could only be real with smaller
amplitude, which is not unusual for Ap stars.  Such a short period, if
confirmed by other authors or in other colors, could be interpreted as the
period of rotation.

\Section{Summary}
In this work we presented a homogeneous series of 63 measurements of the
longitudinal magnetic field $B_e$ in the spectroscopic binary HD~9996,
where the primary companion is an Ap star. Our measurements were obtained
in the years 1994--2011 in the coude focus of the 1-m reflector of the
Special Astrophysical Observatory, using the same spectrometer and analyzer
of circular polarization during 18 years of observations. Using all
available $B_e$ points from 62 years of observations (3 full magnetic
cycles), we improved the long-term magnetic period of HD~9996 and set
$P_{\rm mag}=21.8$~years.

We performed an extensive analysis of line-widths in HD~9996 and
constrained rotational velocity on the equator of the magnetic (primary)
star to $\vv_e\sin i\le8\pm1$~km/s, or its rotational period to $P_{\rm
rot}\ge15\pm2$~days.

Compilation of archival photometric data showed the existence of long-term
variations of HD~9996 consistent with $P_{\rm mag}=21.8$ year. Photometric
data from Hipparcos archive showed the short-term $H_P$ variations with the
period $P_{\rm phot}=39.76$ days, however, the existence of this period was
not confirmed in our spectral analysis of these data. We identify $P_{\rm
mag}$ with the precession period of the primary Ap star.

We regard the binary star HD~9996 as an important object for studies of
generation, evolution and interaction of stellar gravitational and magnetic
fields.

\Acknow{We acknowledge support from Polish Ministry of Science and Higher 
Education grant No. N N203 511638 and Russian grant ``Leading Scientific
Schools'' N5473.2010.2.}

\end{document}